\documentstyle[12pt, aaspp4]{article}
\def\reference{\par\noindent\hangindent 20 pt}

\begin{document}

\input{psfig.tex}

\title{\bf The HST Survey of BL~Lacertae Objects. IV.
Infrared Imaging of Host Galaxies}
 
\author{Riccardo Scarpa, C. Megan Urry, Paolo Padovani\footnote{Also affiliated to the 
Astrophysics Division, Space Science Department, European Space Agency, on
leave from Dipartimento di Fisica, II Universit\`a di Roma ``Tor Vergata''.}, 
Daniela Calzetti \& Matthew O'Dowd} 
\affil{Space Telescope Science Institute}
\authoraddr{Space Telescope Science Institute, 3700 San Martin Dr., 
Baltimore, MD 21218, USA}
\authoremail{scarpa@stsci.edu, cmu@stsci.edu, padovani@stsci.edu,
calzetti@stsci.edu, odowd@stsci.edu}

\begin{abstract}

The HST NICMOS Camera 2 was used for H-band imaging of 12 BL Lacertae objects 
taken from the larger sample observed with the WFPC2 in the R band 
(Urry et al. 2000; Scarpa et al. 2000). 
Ten of the 12 BL Lacs are clearly resolved, and 
the detected host galaxies are large, bright ellipticals with 
average absolute magnitude $\langle M_H \rangle =-26.2\pm0.45$~mag and 
effective radius $\langle r_e \rangle =10\pm 5$~kpc.
The rest-frame integrated color of the host galaxies
is on average $\langle $R-H$ \rangle =2.3\pm0.3$, consistent with
the value for both radio galaxies and normal, non-active elliptical galaxies,
and indicating the dominant stellar population is old.
The host galaxies tend to be bluer in their outer regions than in their cores,
with average color gradient $\Delta (R-H)/\Delta \log r =-0.2$~mag, 
again consistent with results for normal non-active elliptical galaxies. 
The infrared Kormendy relation, derived for the first time for BL Lac host
galaxies, is $\mu_e = 3.8 \log r_e + 14.8$, fully in agreement with 
the relation for normal ellipticals. 
The close similarity between BL Lac host galaxies and normal ellipticals
suggests the active nucleus has surprisingly little effect on the host 
galaxy. This supports a picture in which all elliptical galaxies harbor
black holes which can be actively accreting for some fraction of their
lifetime.

{\underline{\em Subject Headings:}}
BL~Lacertae objects --- galaxies: structure --- galaxies: elliptical

\end{abstract}

\section{Introduction}

The properties of BL Lac host galaxies have been the subject of
intense study. Early work from the ground
found that nearby BL Lacertae objects were surrounded by luminous
elliptical galaxies (Ulrich 1989 and reference therein).
With higher resolution ground-based observations
(Abraham et al. 1991; Stickel et al. 1993; Wurtz, Stocke \& Yee 1996; 
Kotilainen, Falomo \& Scarpa 1998a; Falomo \& Kotilainen 1999),
and now with the 0.1-arcsecond spatial resolution of 
the Hubble Space Telescope (HST)
(Jannuzi, Yanny \& Impey 1997; Yanny, Jannuzi \& Impey 1997;
Falomo et al. 1997; Urry et al. 2000; Scarpa et al. 2000),
this result has been extended to much higher redshifts.
Detection and morphological classification of BL Lac host galaxies 
is now almost routine for redshifts $z\lesssim 0.5$.

Nearly 100 BL Lac host galaxies have been detected so far, 
and there is now abundant evidence that most, and possibly all,
BL Lac host galaxies are ellipticals 
(Abraham et al. 1991; Wurtz et al. 1996; Scarpa et al. 2000).
Although there have been claims of spiral structure 
in a few host galaxies (Abraham et al. 1991; Wurtz et al. 1996),
our HST survey of 110 BL Lac objects 
found overwhelming preference for $r^{1/4}$-law surface brightness profiles,
and no cases where an exponential disk profile was preferred
(Falomo et al. 1997; Urry et al. 2000; Scarpa et al. 2000).
Given this large sample, the fraction of spiral galaxies among BL Lac hosts
must be less than 7\% at the 99\% confidence level (Scarpa et al. 2000).  
This holds for both radio- and X-ray-selected BL Lac objects, and
can be extended (though with less accuracy) to the more powerful
flat spectrum radio quasars (Kotilainen, Falomo \& Scarpa 1998b).

With high-resolution images in multiple filters, it is possible to
probe the stellar populations of AGN host galaxies, and to compare
them to normal ellipticals.
Indeed, since many normal (non-active) galaxies appear to have 
central super-massive black holes 
(Richstone et al. 1998, van der Marel 1999),
perhaps all elliptical galaxies have the potential 
to become active when accretion is turned on.
If so, the stellar populations of AGN host galaxies should be
the same as those of normal ellipticals.
 
Integrated colors and color profiles have been reported for 
a handful of BL Lac host galaxies 
(Urry et al. 1999; Kotilainen, Falomo \& Scarpa 1998a). 
These sparse data suggest that BL Lac host galaxies 
are not strongly affected by the intense activity at their centers,
a result with obvious implications for the galaxy-AGN connection,
and galaxy and AGN evolution.

To further investigate the colors of BL Lac host galaxies, 
we observed 12 BL Lac objects in the H-band with the 
HST\footnote{Based on observations made with the NASA/ESA
Hubble Space Telescope, obtained at the Space Telescope Science 
Institute, which is operated by the Association of Universities 
for Research in Astronomy, Inc., under NASA contract NAS~5-26555.}
Near-Infrared Camera and Multi-Object Spectrometer (NICMOS).
The targets were taken from the larger sample studied previously 
in the R band with the HST Wide Field and Planetary Camera~2 (WFPC2).
For these 12 BL Lac objects we therefore have high-resolution
HST observations in two filters probing rather long wavelengths,
where the bulk of the emission of the old
stellar population lies. These data allow us to derive
color profiles and integrated galaxy colors which 
constrain the stellar content of these host galaxies.
The observed sources cover the redshift range $0.1 < z < 0.5$, 
at the upper end of which cosmic evolution begins to have an effect.

In Section 2 we describe the NICMOS observations and data analysis,
Section 3 gives the results and discusses them,
Section 4 is discussion and conclusions.
Throughout the paper we use H$_0=50$ km/s/Mpc and q$_0=0$.

\section{Observations and Data Reduction}

The HST NICMOS Camera~2 was used to obtain high resolution 
images of 12 BL Lac objects in snapshot mode (that is, during
random gaps in the HST observing schedule). The approved list of 
targets consisted of 29 BL Lac objects from our original WFPC2
snapshot survey list of 132 BL Lacs, comprising seven complete,
flux-limited samples selected in the radio, optical, or X-ray.
The 29 NICMOS targets had existing WFPC2 F702W data, were
uniformly distributed in the redshift range $0.1<z<0.5$, and
include both ``red'' (low-frequency-peaked, high-power) 
and ``blue'' (high-frequency-peaked, low-power) BL Lac jets
(Urry 1999).
In the end, data were obtained for 8 blue and 4 red BL Lacs.
Among them is 0454+844, for which a corrected redshift,
$z>1.3$, was subsequently reported.

Observations were made in the F160W filter, which has 
$\lambda_{eff}=1.6 \mu$m and closely matches the standard H-band.
Camera~2 has pixel size $0.075$~arcsec,
providing at 1.6~$\mu$m Point Spread Function (PSF) sampling as
good as the WFPC2 planetary camera (which has pixel size of $0.046$~arcsec)
in the R-band.
For each BL Lac, we obtained three separate images with increasing 
exposure times, dithered to improve resolution and to better estimate
sky background. The journal of observations is given in Table~1.

Data were first reduced and flux calibrated in the standard HST pipeline. 
The effect of the random bias (known as the ``pedestal'') 
was then removed as follows: 
first, all detected sources in the field of view were excluded;
then, considering each of four quadrants separately, we found
the value of the pedestal that minimized the spread in the remaining 
(non-source) pixels, with the added constraint that the sky background
had to be the same in all four quadrants. 
Corrected images were then cleaned of cosmic
rays and other defects, and a sky frame (obtained from median
filtering the three images) was subtracted. 
To improve the resolution and PSF sampling in the final combined image 
we then re-sampled each image, splitting each pixel into four sub-pixels. 
These re-sampled images were re-centered matching the peak of 
the BL Lac light centroid to within 0.2 pixels, and then combined. 
The final images are presented as contour plots in Figure~1.

The azimuthally averaged, surface brightness radial profiles, shown
in Figure~2, were fitted following the same procedure outlined by 
Scarpa et al. (2000), which is based on a $\chi^2$ minimization strategy. 
To model the PSF we used the Tiny Tim software (Krist 1995). 
For the two unresolved sources in our sample (0851+202 and 0454+884), 
the Tiny Tim PSF model follows the azimuthally averaged
surface brightness profile to at least 3.5~arcsec from the nucleus
(see Fig.~2), showing it correctly describes the true PSF.
In 9 of the other 10 sources, the observed radial profile 
lies significantly above the PSF model, leaving no doubt that the
host galaxy is detected. In the tenth case, 1749+096, the host galaxy
is marginally detected; however, the host galaxy was clearly detected 
in our WFPC2 F702W image (Scarpa et al. 2000) and modeling the 
excess of light above the Tiny Tim PSF we obtain for the host galaxy 
a R-H color consistent with the value 
expected for a normal elliptical galaxy ($R-H=2.3$ at $z=0.32$).
Thus we conclude that 1749+096 is resolved and that 
the Tiny Tim software models the NICMOS Camera~2 PSF very well.

\section{Results and Discussion}

\subsection{Morphology and Luminosity of the Host Galaxies}

Snapshot images are typically shorter than would be ideal
for studying extended emission from the host galaxy. Therefore rather than 
using a full two dimensional approach, we
chose to extract azimuthally averaged radial 
profiles, which can be traced out to $\sim$3--4~arcsec 
from the center (Fig.~2), significantly increasing the signal-to-noise and 
providing well-constrained total magnitude and effective radius
for the host galaxies, at the expenses of some spatial information like
ellipticity and position angle. 
From the two-dimensional analysis of the R-band data of
all $z<0.2$  sources observed in the optical survey, which 
included 5 of the 12 objects presented here, 
we know that BL Lac host galaxies are quite round and have undisturbed
morphology (Falomo et al. 2000). Thus, we are confident that our one-dimensional 
approach is not biased or affected by unusual feature of the host galaxy. 

Ten of the 12 observed BL Lac objects are clearly resolved in the 
NICMOS images.
The two unresolved BL Lacs are the distant source 0454+844 ($z>1.3$), 
and 0851+202 (OJ~287, $z=0.306$), which has a very luminous nucleus 
and has so far resisted all attempts to resolve its host galaxy
(see Yanny, Jannuzi \& Impey 1997 and counter 
arguments by Sillanp\"a\"a et al. 1998).

The average radial profiles are in all cases well described by a
PSF plus a de Vaucouleurs ($r^{1/4}$) law (Figure~2). 
In no case is a disk galaxy preferred (for 0502+675 and 1749+096 
the two galaxy models are equally acceptable.) 
The average extinction and K-corrected absolute H magnitude 
of the 10 resolved hosts is
$\langle M_H \rangle =-26.2\pm0.45$~mag (standard deviation). 

The average effective radius is $\langle r_e \rangle =10\pm 5$~kpc, 
and it ranges from 3 to 20~kpc. 
The derived magnitudes and effective radii are given in Table~2.
These results fully confirm what was found from the R-band HST images,
that the host galaxies are giant ellipticals,
on average $\sim1$~mag brighter than an $M^*$ galaxy
($M_H^*=-25.0\pm 0.3$ galaxy; Mobasher, Sharples \& Ellis 1993).

\subsection{Host Galaxy Integrated Color}

The extinction-corrected host galaxy total magnitudes, derived from the 
best fit de Vaucouleurs model allow us to derive integrated 
R--H colors for all 10 resolved sources (R-band data are in
Table~2 of Urry et al. 2000). Results, summarized in Table 3, show
that the average BL Lac host galaxy color is R--H=$2.3\pm0.3$ mag,
consistent with the value $2.2\pm 0.5$ found for 5 more BL~Lac objects by 
Kotilainen, Falomo \& Scarpa (1998a), as well as with 
the value reported for a sample 
of 12 nearby normal elliptical galaxies, R--H=$2.3\pm 0.2$~mag 
(Peletier et al. 1990; value converted from V--K assuming
V--R=0.58 and H--K=0.22, according to Recilla-Cruz et al. 1990). 

In the top panel of Figure~3 we plot the R--H 
colors of our BL Lac host galaxies, computed from extinction and 
K-corrected magnitudes derived from our WFPC2 and NICMOS data.
(In the absence of evolution, the K correction removes the
dependence of color on redshift).
The BL Lac host galaxy colors cluster around the average color
of normal, $z=0$ galaxies (dashed line in Fig.~3).

We now compare the BL Lac host galaxies to radio galaxies for which
infrared colors are available.
From the work of Lilly and Longair (1982),
eight 3C radio galaxies (3C33, 98, 123, 192, 219, 295, 299, and 388) 
have measured R--K colors, $z<0.5$ (as in our BL Lac sample), 
and are not classified as N-galaxies (whose color is affected by the 
light from the bright nucleus).
After K-correcting the R-band data (the correction is negligible
for the K band), and assuming H--K = 0.22 (Recilla-Cruz et al. 1990), we obtain
$\langle R-H \rangle =2.3 \pm 0.3$~mag, fully consistent with what we found
for BL Lac host galaxies. 
It is true that the radio galaxies in Lilly \& Longair (1982) are mostly FR~II
(high-luminosity) sources, while BL Lac objects are most likely associated
with FR~I (low-luminosity) radio galaxies (Urry \& Padovani 1995);
however, there is growing evidence that FR~I and FR~II host galaxies 
have similar optical properties (Govoni et al. 2000; Scarpa \& Urry 2000),
so this is unlikely to distort the comparison.

Similar colors were reported by De Vries et al. (1998) for a 
different sample of radio galaxies of various morphological type,
at $\langle z \rangle \sim0.3$.
They found $\langle R-H \rangle = 2.4$~mag
(transformed from observed $\langle R-K \rangle = 3.0$~mag 
assuming H-K=0.22~mag and K-cor=0.4~mag for the R band; Fukugita et al. 1995), 
again consistent with the value for BL Lac hosts. 

The R--H color depends on the age of the dominant stellar
population in these galaxies. Due to the metallicity-age degeneracy,
the limited information provided by a single color does not allow us
to set a tight limit on the epoch of major star formation in these
galaxies; however, our data are consistent with the expectation for a
single, very old, coeval stellar population. In the bottom panel in
Figure~3, the BL Lac host galaxy colors (now not K corrected) are
shown to evolve similarly to stellar synthesis models with ages 
of formation of 4 and 10 Gyrs (solid lines). Similar results have been
found for radio galaxies and non-active ellipticals (McCarthy 1993).

\subsection{Radial Color Gradients}

We derived R--H color profiles for the ten resolved sources 
combining NICMOS H-band data with the WFPC2 R-band data from 
Scarpa et al. (2000). Two of the ten (0502+675 and 1749+096)
had such noisy profiles, due to
the weakness of the host galaxy relative to the central point source,
that we do not discuss them further.
The eight remaining color profiles are shown in Figure~4. 
All but one host galaxy show a clear trend for R--H to decrease 
with increasing radius, that is, the galaxies become bluer in 
the outer regions. 
Table~3 gives for each object the slope of the color profile
in the (R--H) - Log(r) plane. 

Similar R--H color gradients for five BL Lac host galaxies were found by 
Kotilainen, Falomo \& Scarpa et al. (1998a).
Urry et al. (1999) reported V--I color profiles for six BL Lac objects
based on HST WFPC2 data; because of the smaller wavelength baseline, 
any color gradients are weaker than in the present R--H color profiles,
but the blue-ing trend is still visible in at least
three host galaxies (1407+595, 2143-070, and 2254+074).

To better quantify this color gradient, we constructed an average
R--H color profile, corresponding to the implicit assumption that
all host galaxies have the same gradient (a likely oversimplification).
We first normalized each profile by subtracting its average color, 
then averaged the eight normalized profiles, as shown in Figure~5. 
The error bars represent the standard deviation of the data 
in each 1-kpc interval, and thus reflect the variation in color from
object to object rather than measurement errors.
The best-fit linear color gradient has slope 
$\Delta (R-H)/\Delta \log \ r =-0.2\pm 0.1$~mag,
consistent with previous reports
($\Delta (R-H)/\Delta \log \ r =-0.09\pm 0.04$~mag for
the five BL Lacs in Kotilainen, Falomo \& Scarpa 1998a). 

The present data offer strong evidence that BL Lac host galaxies 
become bluer away from the galaxy center. An immediate consequence 
(and a consistency check) is that the effective radius of the 
galaxy is a function of the selected filter.
In Figure~6 we compare the best-fit effective radii derived from
our HST data, i.e., NICMOS for the H band and WFPC2 for the R band.
Even though errors are in some cases large, there is a clear tendency 
for the effective radius to be larger in the optical than in the infrared.
This is clearly because as wavelength decreases a larger fraction of the 
total light lies in the external regions, i.e., the host galaxies are
bluer on their outskirts.

Color gradients of the same sign have been found in several studies
of normal elliptical galaxies
(Boroson et al. 1983; 1987; Cohen 1986; Peletier et al.  1990; Munn 1992).
A value of $\Delta (V-K)/\Delta \log \ r = -0.16\pm 0.18$~mag was found by
Peletier et al. (1990) for a sample of 12 elliptical galaxies.
Given the median redshift of our BL Lac sample ($z=0.2$), 
our R-band profiles are roughly equivalent to their V-band data at $z=0$;
therefore, assuming H and K profiles are similar 
(since the red stars evolve only slowly), 
the V--K gradient measured by Peletier et al. is roughly equivalent to 
our R--H color gradient.
The fact that our two values are similar therefore
underscores the similarity of BL Lac host galaxies and normal ellipticals.
Specifically, the stellar populations of active and non-active 
ellipticals must not be very different, not only in an integrated sense 
but differentially across the galaxy.

\subsection{The Surface Brightness -- Effective Radius Relation}

Infrared observations map the bulk of the slowly evolving stellar populations
of galaxies, and are therefore particularly appropriate for investigating
luminosity and galaxy morphology. It is therefore useful to investigate
the Kormendy relation between effective radius, $r_e$, and 
surface brightness at that radius, $\mu_e$, in the H band.
Plotting our best-fit values for $\mu_e$ and $r_e$ for BL Lac host
galaxies (Table~2) shows they follow the same
well-defined correlation as non-active elliptical galaxies (Figure~7).
The dispersion around the best linear fit is
comparable to that found in the R band (Urry et al. 2000).
Since the H-band fits are less well constrained (the uncertainties
are larger; see Figure~6), the intrinsic correlation must be tighter, 
reinforcing that infrared observations are indeed 
optimal for investigating the fundamental plane of ellipticals.

A linear fit to our data combined with those from Kotilainen, Falomo \& 
Scarpa (1998a) 
gives $\mu_e = 3.8\log r_e +14.8$~mag/arcsec$^2$, where $r_e$ is in 
kiloparsecs. In the K band, for a sample of 59 normal elliptical galaxies,
$\mu_e = 4.3\log r_e +14.3$~mag/arcsec$^2$ has been found (Pahre et al. 1995, 
after conversion to our adopted cosmology). The agreement is very
good considering the restricted range in $r_e$ spanned by BL Lac 
hosts, which are all giant galaxies. Again, this result fully
supports the picture in which all elliptical galaxies 
have the potential to experience a phase of nuclear activity.

\section{Summary and Conclusions}

The HST NICMOS Camera~2 was used to image 12 BL Lac objects in the H
band.  All sources were also observed previously with the HST WFPC2 in
the R band.  Ten of the 12 BL Lacs were clearly resolved in the H
band, with the remaining two having luminous nuclei that probably
swamp any surrounding galaxy light.  From direct inspection of the
NICMOS data (Figure 1), as well as from the R-band WFPC2 data (Falomo
et al. 2000), we found no indication of distortion, asymmetry, or
tails in the host galaxies, down to surface brightness $\mu_H\sim18.5$
mag/asec$^2$. To increase the signal-to-noise ratio and reach fainter
structure, we extract for each source the azimuthally averaged radial
profile, which is traced out to 3-4 arcsecs from the nucleus down to
$\mu_H\sim21$ mag/asec$^2$ (Figure 2). This corresponds to 1-3
effective radii depending on galaxy distance and size.

The ten detected host galaxies are luminous ellipticals, with
$r^{1/4}$-law profiles. 
Their average K-corrected absolute H magnitude is
$\langle M_H \rangle =-26.2$, with $1\sigma$ dispersion of $0.45$~mag, 
about one magnitude brighter than an M$^*$ galaxy
($M_H^*=-25.0\pm 0.3$ galaxy; Mobasher, Sharples \& Ellis 1993),
and their average effective radius is $\langle r_e \rangle =10\pm 5$~kpc.

The average R--H color of the BL Lac host galaxies,
$\langle R-H \rangle = 2.2 \pm 0.3$~mag,
is typical of normal ellipticals or radio galaxies, and indicates
the dominant stellar population is more than a few gigayears old.
Because of the high resolution of HST, we were able to detect color
gradients in the majority of BL Lac host galaxies, such that their
cores are redder than their external regions. From our color profiles,
which probe a region extending from 0.2 to 1.5 effective radii,
we measure an average color 
gradient of $\Delta (R-H)/\Delta \log \ r =-0.2\pm 0.1$~mag, again 
similar to those observed in elliptical galaxies (Peletier et al. 1990).

The infrared Kormendy relation for BL Lac host galaxies is 
indistinguishable from the same relation for normal elliptical
galaxies. Thus they appear to occupy the same region of the
fundamental plane as other ellipticals, including radio galaxies.

In every respect we could measure, the host galaxies of BL Lac objects
appear to be normal ellipticals.  Given the high precision of the HST
data, our study was largely free of confusion from the bright nucleus.
Had there been dust lanes or other morphological peculiarities beyond
$\sim 0.2$~arcsec from the center, they would have been readily
apparent.  We conclude therefore that to first order, the host galaxy
``knows'' nothing about the active nucleus it harbors.  This is in
contrast to claims of a nuclear-host connection in more powerful AGN
(McLure et al. 1999, McLeod \& Rieke 1995).  This result argues
against AGN and galaxies being separate entities, and instead supports
a picture in which all galaxies can be AGN but with a relatively low
duty cycle. This implies all galaxies (at least, elliptical galaxies)
have super-massive black holes at their centers, as indeed appears to
be the case in many nearby ellipticals (Richstone et al. 1998, van der
Marel 1999).

Because active galaxies are often observed in clusters, it has been
proposed that galaxy mergers are responsible for initiating the
nuclear activity. In the HST images of BL Lac objects we do not see
any of the morphological peculiarities (e.g., tails, dust lanes) or
unusual stellar content which might be expected following a merger.  A
possibility is that in the vast majority of cases the nuclear activity
is triggered by merge of a companion galaxy so small that the global
properties of the main galaxy remain unchanged.  However, better
constraints on this hypothesis would have been possible were we to
obtain additional images in a bluer filter.  In any case, the present
work strongly supports a picture in which all galaxies have the
potential to be active.  This in turn suggests the formation and
evolution of galaxies occurred hand in hand with the growth of
super-massive black holes at their centers.

\acknowledgements
Support for this work was provided by NASA through grant number
GO-07893.01-96A from the Space Telescope Science Institute, which is
operated by AURA, Inc., under NASA contract NAS~5-26555.
This research made use of 
NASA's Astrophysics Data System Abstract Service (ADS).

\noindent {\bf References}

\reference Abraham R.G., McHardy I.M. \& Crawford C.S. 1991, MNRAS 252, 482
\reference Boroson T.A., Thompson I.B. \&  Shectman S.A. 1983, AJ 88, 1707
\reference Boroson T.A. \& Thompson I.B. 1987, AJ 93, 33
\reference Cohen J.G. 1986, AJ 92, 1039
\reference de Vries W.H., O'Dea C.P., Baum S.A. \& Perlman E. 1998, ApJ 503, 156
\reference Falomo R. \& Kotilainen J.K. 1999, A\&A 352, 85 
\reference Falomo R., Urry C.M., Pesce J.E., Scarpa R., Giavalisco M. \& 
	Treves A. 1997, ApJ 476, 113
\reference Fukugita M., Shimasaku K. \& Ichikawa T. 1995, PASP 107, 945
\reference Govoni F., Falomo R., Fasano G. \& Scarpa R. 2000, A\&A 353, 507 
\reference Jannuzi B.T., Yanny B. \& Impey C. 1997, ApJ 491, 146
\reference Kotilainen J.K., Falomo R. \& Scarpa R. 1998a, A\&A 336, 479
\reference Kotilainen J.K., Falomo R. \& Scarpa R. 1998b, A\&A 332, 503
\reference Krist, J. 1995, in Astronomical Data Analysis, Software
	and Systems IV, eds. R. Shaw et al., 
	(San Francisco: Astr. Soc. Pac.), p. 349
\reference Lilly S.J. \& Longair M.S. 1982, MNRAS 199, 1053
\reference McCarthy P. 1993, PASP 105, 1051
\reference McLeod K.K. \& Rieke G.H. 1995, AJ 109, 1979
\reference McLure R.J., Kukula M.J., Dunlop J.S., Baum S.A., O'Dea C.P. \&
	Hughes D.H. 1999, MNRAS 308, 377
\reference Mobasher B., Sharples R. M. \& Ellis R. S. 1993, MNRAS 263, 560
\reference Munn J.A. 1992, ApJ 399, 444
\reference Pahre M.A., Djorgovski S.G. \& De Carvalho R.R. 1995, ApJL 435, 17
\reference Peletier R.F., Valentijn E.A. \& Jameson R.F. 1990, A\&A 233, 62
\reference Recillas-Cruz E., Carrasco L., Serrano A. \& Cruz-Gonzales I. 1990,
	A\&A 229, 64
\reference Richstone, D., et al. 1998, Nature 395, 14
\reference Scarpa R. \& Urry C. M. 2000, in preparation
\reference Scarpa R., Urry C. M., Falomo R., Pesce J. E. \& Treves A. 2000, 
	ApJS, 532, 740
\reference Sillanp\"a\"a A., Takalo L.O., Nilsson K., Pursimo T. \&
	Pietil\"a 1998, in {\it BL Lac Phenomenon}, A.S.P. Conf. Ser., 
	Vol. 159, ed. L.T. Takalo and A. Sillanp\"a\"a, p. 395
\reference Stickel M., Fried J.W., \& K\"uhr H. 1993, A\&AS 98, 393
\reference Urry C.M. \& Padovani P. 1995, PASP 107, 803
\reference Urry C.M., Scarpa R., O'Dowd M., Falomo R., Pesce J.E., 
	Treves A. \& Giavalisco M. 1999, ApJ 512, 88
\reference Urry C.M., Scarpa R., O'Dowd M., Falomo R., Pesce J.E., 
	\& Treves A. 2000, ApJ, 532, 861
\reference Urry, C.M. 1999 Astroparticle Physics 11, 159
\reference Ulrich M.-H. 1989, in ``BL Lac objects', ed. L. Maraschi, T. 
	Maccacaro and M.-H. Ulrich (Springer, Berlin), p. 92
\reference van der Marel, R. 1999, AJ 117, 744
\reference Wurtz R., Stocke J.T. \& Yee H.K.C. 1996, ApJS 103, 109
\reference Yanny B., Jannuzi B.T. \& Impey C. 1997, ApJ 484, L113

\newpage

{\bf Figure Caption.}

{\bf Figure 1:}
Contour plots from the central $7.5 \times 7.5$~arcsec regions of the
combined NICMOS F160W images of all 12 BL Lac objects.
Isophotes are spaced by 0.5 magnitudes and the surface brightness of the fainter
is given in each panel (in mag/arcsec$^2$). To improve the visibility of
low surface brightness structure, the images
have been smoothed using a Gaussian filter with width 0.0375~arcsec.
The arrow points north.
The host galaxies are resolved in 10 of 12 cases, the two unresolved 
BL Lacs being 0454+884 and 0851+302. Whatever structure is visible in 
the latter two images is due to the point spread function, as is
the bridge that appears to connect 0607+710 to the nearby star.

{\bf Figure 2:}
Azimuthally averaged  surface brightness radial profile ({\it points with 
error bars}), and the best-fit point source plus de Vaucouleurs model 
(or only the point source for unresolved sources). 
{\it Dotted line:} the point source model (PSF). 
{\it Dashed line:} de Vaucouleurs model convolved with the PSF. 
{\it Solid line:} sum of the two model components. 
When not visible, the dashed line is superimposed on the solid line, and the
solid line is covered by the observed points.
The PSF model works very well, as evidenced by how it fits the profiles
of the unresolved BL Lacs (0454+884 and 0851+202), out to more than
3~arcsec. Nine of the remaining 10 objects have strong excess emission above
the PSF; the tenth, 1749+096, has a host galaxy only marginally detected 
above the PSF but at the level expected given the HST R-band detection.

{\bf Figure 3:}
BL Lac host galaxy R--H color as a function of redshift,
showing their similarity to normal ellipticals with old stellar populations.
{\it Solid squares:} our HST data; 
{\it open squares:} ground-based data from Kotilainen, Falomo \& 
Scarpa (1998a).  
{\bf Upper panel:} Rest-frame colors, with both extinction and 
K corrections applied. The horizontal line at R--H=2.3~mag 
is the average color observed for normal elliptical galaxies
(Kotilainen Falomo \& Scarpa 1998a).
{\bf Lower panel:} Observed colors (i.e., extinction- but not K-corrected) 
of BL Lac host galaxies compared to the expected colors for coeval stellar
populations with age 10 Gyrs {\it (upper line)} and 4 Gyrs {\it (lower line)}.

{\bf Figure 4:}
R-H color profiles of BL Lac host galaxies. 
Two objects, 0502+675 and 1749+096, are omitted because the
signal-to-noise ratios for their host galaxies are very low.
With the exception of 0229+200,
the host galaxies become significantly bluer in their
outer regions, as observed in normal elliptical galaxies
(Peletier et al. 1990).

{\bf Figure 5:}
Average color profile for the 8 well-resolved sources, derived from
profiles in Figure~4. Profiles were combined after first subtracting 
their average color, then computing the weighed average of the points 
on intervals of 1~kpc. The formal statistical error on each point 
is smaller than the
size of the symbol, while the intrinsic dispersion of the points in
each bin is shown at the top-left corner of the figure.
The best fit slope {\it (solid line)} is 
$\Delta (R-H)/\Delta \log \ r =-0.2\pm 0.1$~mag.

{\bf Figure 6:}
Comparison of effective radii derived from WFPC2 (R-band) and 
NICMOS (H-band) images. The solid line is the locus of an exact match 
between the two measures,
while the dashed line is a linear fit to the data.
This shows that the R-band effective radius is larger than the H-band
one, reflecting the fact that the host galaxies become 
systematically bluer in their outer regions. 
Two objects show an opposite behavior;
for 1749+096, the measurement has large uncertainties, 
while for 0229+200 the lack of color gradient may be real.

{\bf Figure 7:}
The H-band $\mu_e - r_e$ relation for BL Lac 
host galaxies. 
{\it Solid squares:} our HST data;
{\it open squares:} ground-based data from Kotilainen, Falomo \& Scarpa (1998a).
{\it Solid line:} linear fit to the data; 
{\it dashed line:} best-fit relation for a sample of 59 
normal ellipticals reported by Pahre et al. (1995).

\newpage

\begin{deluxetable}{llcr}
\tablenum{1}
\tablewidth{4in}
\tablecaption{{\bf Journal of the Observations}}
\tablehead{
 \colhead{BL Lac} & \colhead{Sample/}  & \colhead{Date} & 
     \colhead{Exp.Time$^{(a)}$} \\ 
\colhead{Object}  & \colhead{Name} &  \colhead{Obs.} &\colhead{(sec)} 
}
\startdata
0229+200 & 1ES      & 06-Jan-98 &   320  \nl
0454+844 & S5       & 26-Sep-98 &  1920  \nl
0502+675 & 1ES      & 04-May-98 &  1428  \nl
0607+710 & MS       & 03-Jun-98 &   896  \nl
0706+591 & EXO      & 20-Apr-98 &   352  \nl
0851+202 & 1Jy/OJ287& 18-Oct-98 &   896  \nl
1212+078 & 1ES      & 10-Mar-98 &   352  \nl
1407+595 & MS       & 20-Sep-98 &  1408  \nl
1418+546 & PG       & 25-Sep-98 &   352  \nl
1749+096 & PKS      & 09-Aug-98 &  1024  \nl
2143+070 & EMSS     & 09-Nov-98 &   960  \nl
2356-309 & HEAO-A3  & 14-Oct-98 &   831  \nl
\enddata
\tablenotetext{(a)}{Total exposure time, i.e., sum of exposure times 
for all images.}
\end{deluxetable}

\newpage

\begin{center}
\scriptsize
\hoffset -5cm
\begin{deluxetable}{cccccccccccc}
\tablenum{2}
\tablewidth{6.7in}
\tablecaption{{\bf Host Galaxy Properties}}
\tablehead{
\colhead{BL Lac}   & \colhead{z} & \colhead{A$_H^{(a)}$} & \colhead{K Corr.} &\colhead{$m_H^{(b)}$} & \colhead{$m_H^{(c)}$} & 
\colhead{$m_H^{(d)}$} & \colhead{$\mu_e^{(e)}$} & \colhead{$r_e^{(f)}$} & \colhead{$M_H^{(g)}$} &  
\colhead{M$_H^{(h)}$} & \colhead{$r_e$}  \\ 
\colhead{object}  &  & & \colhead{H band} &\colhead{(tot)} & \colhead{(PSF)} & \colhead{(host)} & 
& \colhead{(arcsec)} & \colhead{(PSF)} & \colhead{(host)}  & \colhead{(kpc)}  
}
\startdata
0229+200 & 0.139 & 0.13 & 0.01& 13.56 &$ 16.0\pm0.2 $&$ 12.8\pm0.1 $&$ 18.38 $&$ 3.8\pm  0.6$&$ -23.89 $&$ -27.1 $&$ 12 \pm  2  $ \nl
0454+844 &$>$1.34& 0.09 & ... & 16.60 &$    ...     $&$  ...       $&$ ...   $&$    ...     $&$ -29.13 $&$  ...  $&$    ...     $ \nl
0502+675 & 0.314 & 0.16 & 0.03& 18.56 &$ 18.7\pm0.2 $&$ 15.7\pm0.3 $&$ 16.20 $&$ 0.5\pm  0.2$&$ -23.16 $&$ -26.1 $&$ 3.0\pm 1.2 $ \nl
0607+710 & 0.267 & 0.13 & 0.02& 15.05 &$ 16.8\pm0.2 $&$ 14.8\pm0.1 $&$ 18.40 $&$ 1.9\pm  0.3$&$ -24.63 $&$ -26.6 $&$10.5\pm 1.6 $ \nl
0706+591 & 0.125 & 0.08 & 0.01& 13.91 &$ 15.7\pm0.1 $&$ 13.4\pm0.1 $&$ 18.34 $&$ 2.7\pm  0.2$&$ -23.89 $&$ -26.2 $&$ 8.2\pm 0.6 $ \nl
0851+202 & 0.306 & 0.05 & 0.03& 16.60 &$ ...        $&$  ...       $&$ ...   $&$    ...     $&$ -25.08 $&$  ...  $&$    ...     $ \nl
1212+078 & 0.136 & 0.02 & 0.01& 13.92 &$ 15.8\pm0.2 $&$ 13.6\pm0.1 $&$ 17.91 $&$   2\pm  0.6$&$ -23.93 $&$ -26.1 $&$ 6.5\pm 1.9 $ \nl
1407+595 & 0.495 & 0.02 & 0.11& 16.05 &$ 17.0\pm0.2 $&$ 16.1\pm0.2 $&$ 18.76 $&$ 1.7\pm  0.3$&$ -25.87 $&$ -26.9 $&$13.7\pm 2.4 $ \nl
1418+546 & 0.152 & 0.01 & 0.01& 12.74 &$ 13.0\pm0.1 $&$ 13.7\pm0.2 $&$ 18.84 $&$   3\pm  0.3$&$ -26.97 $&$ -26.3 $&$10.7\pm 1.1 $ \nl
1749+096 & 0.320 & 0.06 & 0.03& 13.15 &$ 13.2\pm0.1 $&$ 15.3\pm0.3 $&$ 19.98 $&$ 3.3\pm  1.5$&$ -28.60 $&$ -26.5 $&$20  \pm 9   $ \nl
2143+070 & 0.237 & 0.07 & 0.02& 15.22 &$ 16.2\pm0.2 $&$ 15.3\pm0.2 $&$ 18.55 $&$ 1.5\pm  0.4$&$ -24.88 $&$ -25.7 $&$ 7.6\pm 2.0 $ \nl
2356--309& 0.165 & 0.02 & 0.01& 14.41 &$ 15.2\pm0.1 $&$ 14.7\pm0.1 $&$ 18.54 $&$ 1.7\pm  0.2$&$ -24.97 $&$ -25.5 $&$ 6.5\pm 0.8 $ \nl
\hline												         	       	  
Median   & 0.237 &      &     &       &              &      14.75   &         &              &          &$ -26.2 $&$    9.4     $ \nl
\enddata
\tablenotetext{(a)}{Interstellar extinction in magnitudes, derived assuning $A_H = 0.25 A_R$}
\tablenotetext{(b)}{Observed H-band magnitude of the BL Lac source, 
	integrated to the last point of the radial profile shown in Figure~2.
	Values in this column are neither extinction nor k corrected.
	Systematic+statistical errors for these values are $\sim 10\% $.}
\tablenotetext{(c)}{H-band magnitude of the point source from the best-fit PSF, neither extintion nor k corrected.}
\tablenotetext{(d)}{H-band total magnitude of the host galaxy, obtained 
	integrating the best-fit de~Vaucouleurs law to infinite radius. No extinction nor k correction applied.} 
\tablenotetext{(e)}{H-band surface brightness of the host galaxy at the 
	effective radius, $r_e$, in mag/arcsec$^2$. Values include extinction correction, K-correction, 
	and cosmological dimming.} 
\tablenotetext{(f)}{Effective radius of the best-fit de~Vaucouleurs law, 
	i.e., radius which encloses half the total light.}
\tablenotetext{(g)}{Absolute magnitude of the central point sources, computed for 
	H$_0=50$ km/s/Mpc, q$_0=0$, and corrected for interstellar 
        extinction. K-correction is assumed to be zero because BL Lac nuclei have power law spectra with
        spectral index $\sim -1$.}
\tablenotetext{(h)}{Absolute total magnitudes of the host galaxy, extinction and k-corrected.}
\end{deluxetable}
\normalsize
\hoffset 0cm
\end{center}

\newpage

\begin{deluxetable}{lcc}
\tablenum{3}
\tablewidth{3.5in}
\tablecaption{{\bf Host Galaxy Colors}}
\tablehead{
 \colhead{Object}  & \colhead{R--H$^{(a)}$} & 
\colhead{$\Delta$ (R--H)$/\Delta log(r)$} \\ 
 & \colhead{(mag)} & \colhead{(mag)} \\
}
\startdata
0229+200 &$ 2.5\pm 0.1 $&$ +0.0$ \nl
0502+675 &$ 2.3\pm 0.3 $&$  ...$ \nl     
0607+710 &$ 2.3\pm 0.1 $&$ -0.3$ \nl
0706+591 &$ 2.2\pm 0.1 $&$ -0.3$ \nl
1212+078 &$ 2.2\pm 0.1 $&$ -0.6$ \nl
1407+595 &$ 2.1\pm 0.2 $&$ -0.8$ \nl
1418+546 &$ 2.2\pm 0.2 $&$ -0.5$ \nl
1749+096 &$ 2.9\pm 1.3 $&$ ... $ \nl
2143+070 &$ 1.7\pm 0.2 $&$ -0.6$ \nl
2356-309 &$ 2.2\pm 0.1 $&$ -0.3$ \nl
\hline
Mean Profile &$        $&$ -0.2$ \nl
\enddata
\tablenotetext{(a)}{Integrated rest frame R--H color, K-corrected and
	corrected for extinction. Color gradients are expressed in 
	magnitudes per decade, i.e., it is the variation of color
        for $\Delta log(r)=1$.}
\end{deluxetable}

\newpage

\begin{figure}
\psfig{file=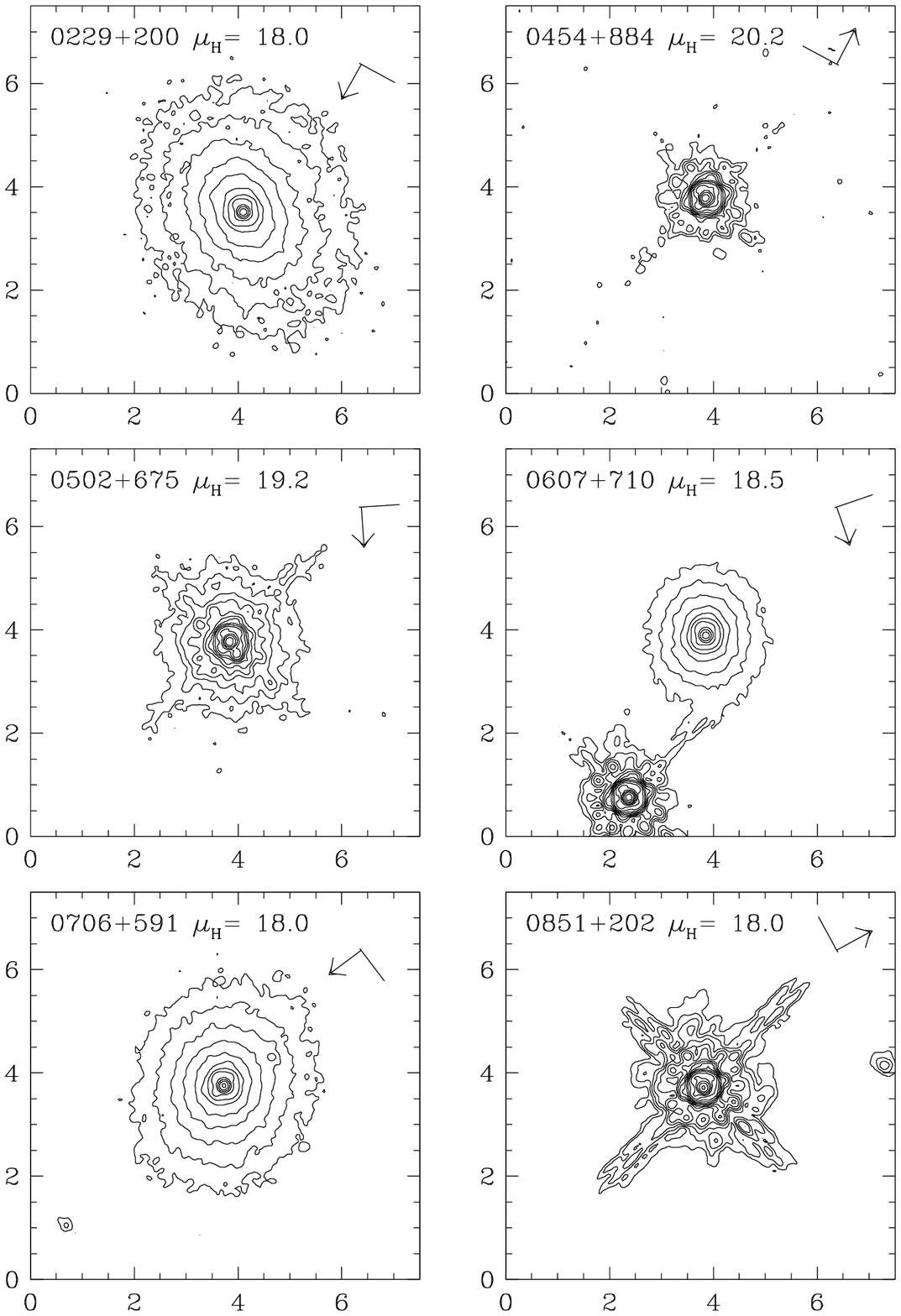,width=14cm}
\caption{ 
}
\end{figure}

\begin{figure}
\psfig{file=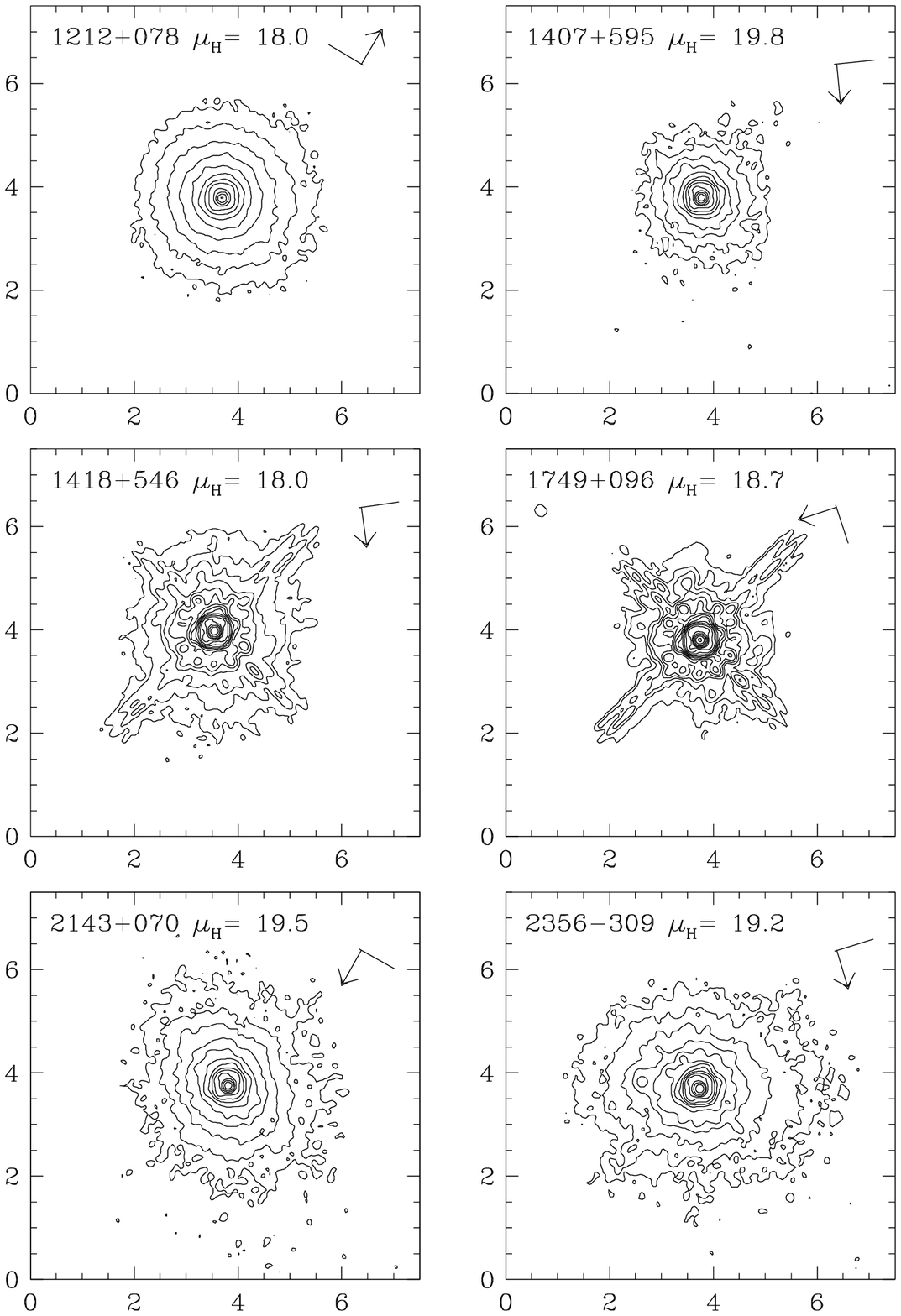,width=14cm}
\end{figure}

\begin{figure}
\vspace{-1cm}
\psfig{file=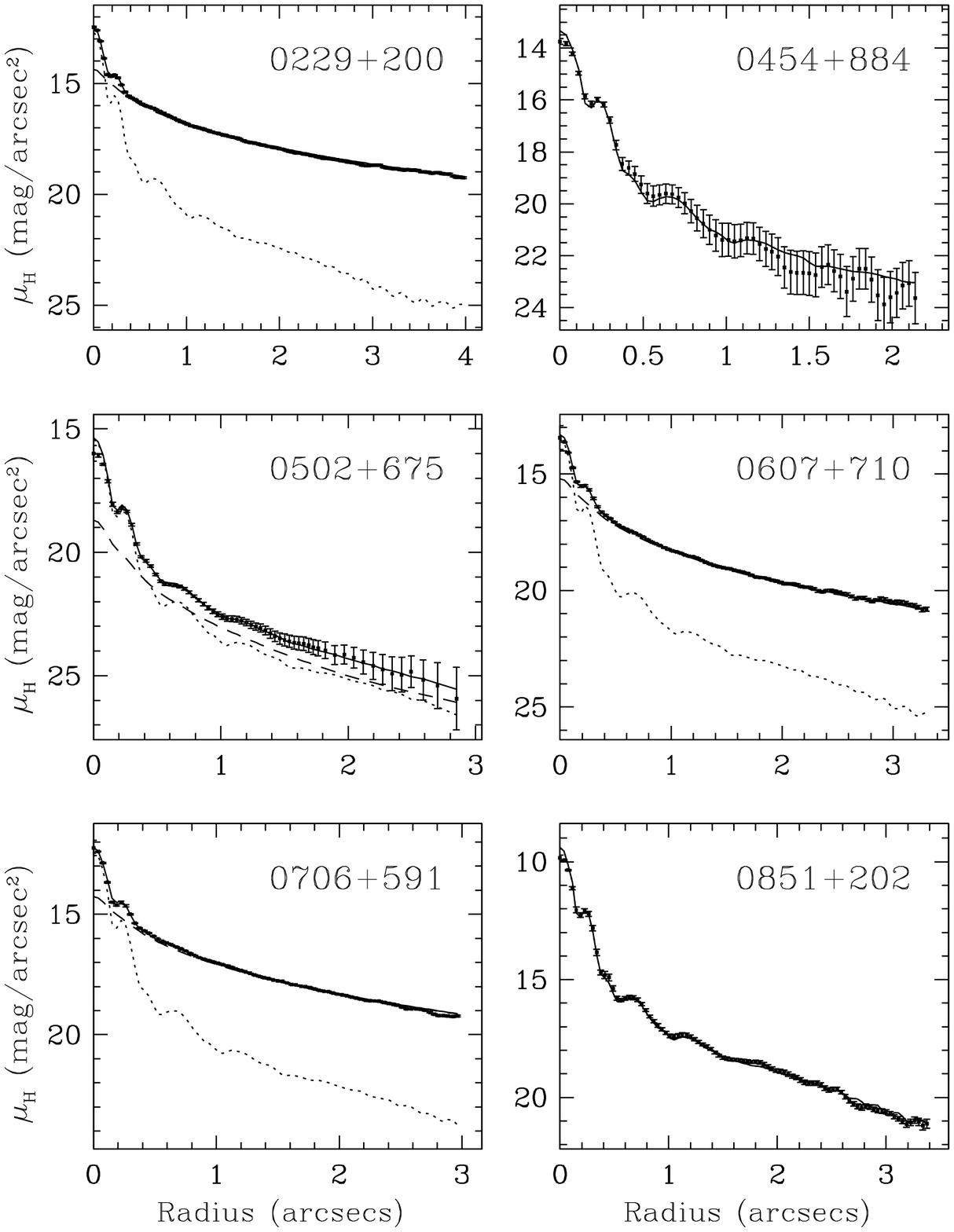,width=14cm}
\caption{
}
\end{figure}

\begin{figure}
\psfig{file=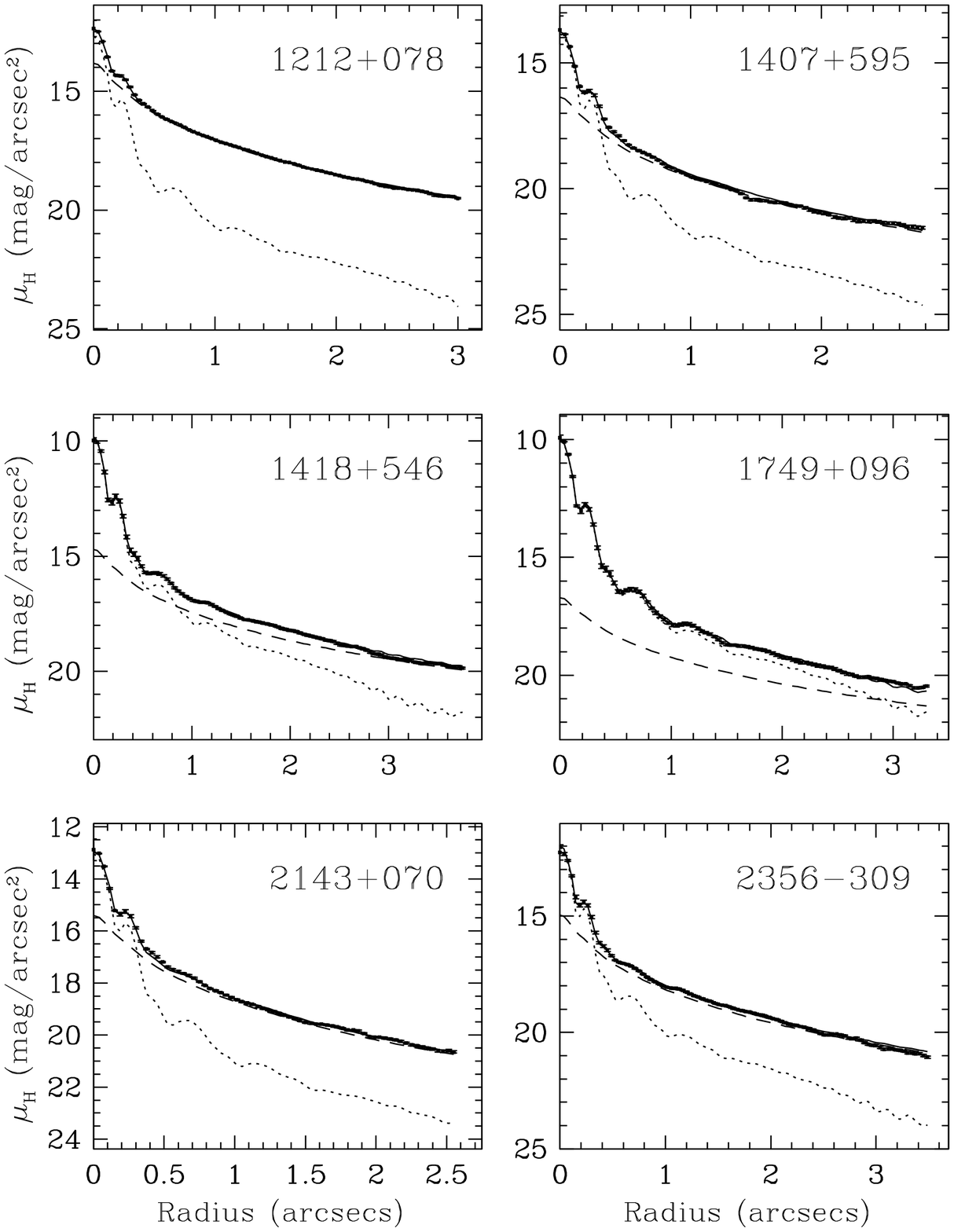,width=14cm}
\end{figure}

\begin{figure}
\psfig{file=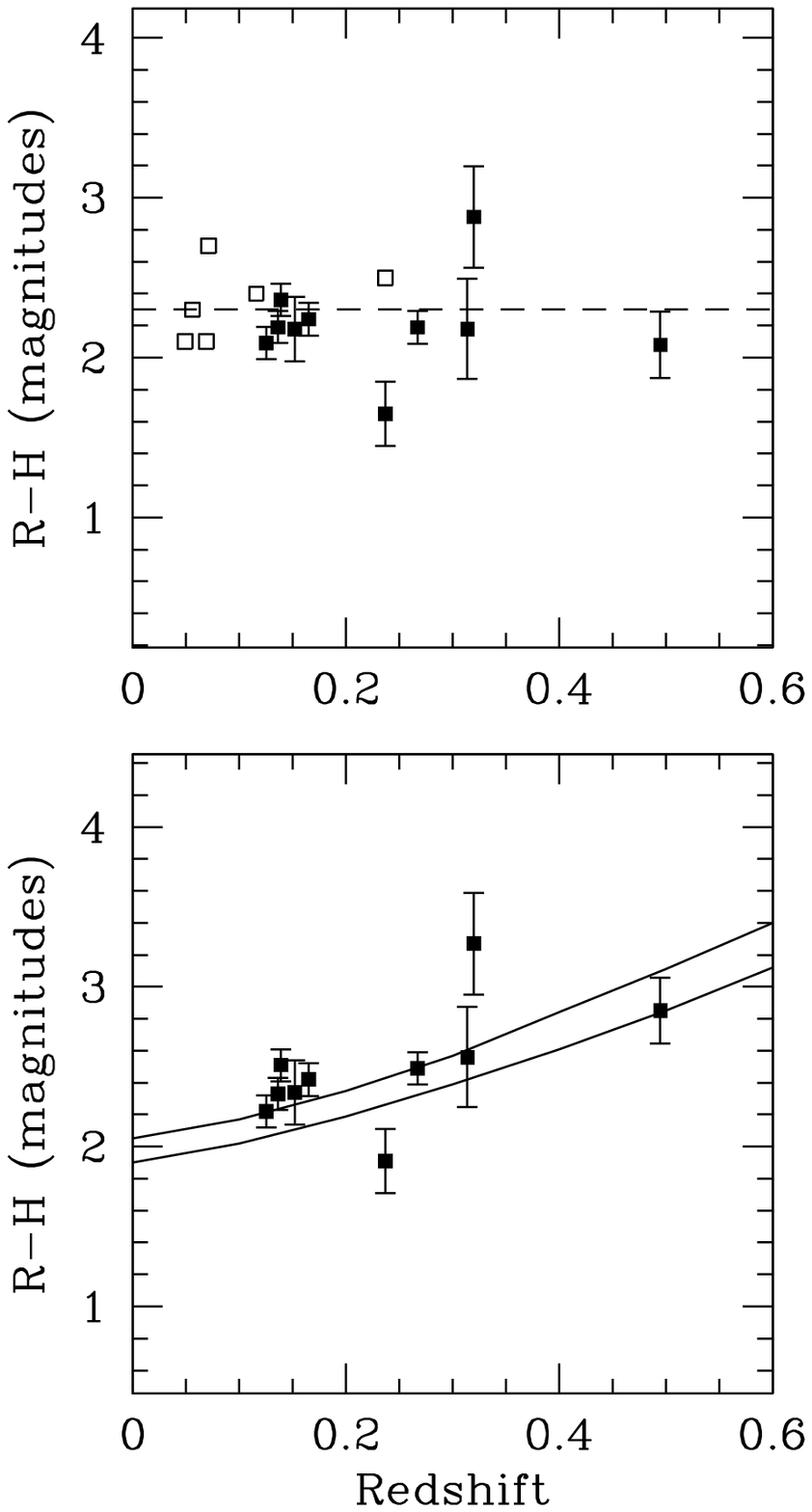,width=11cm}
\caption{
}
\end{figure}

\begin{figure}
\psfig{file=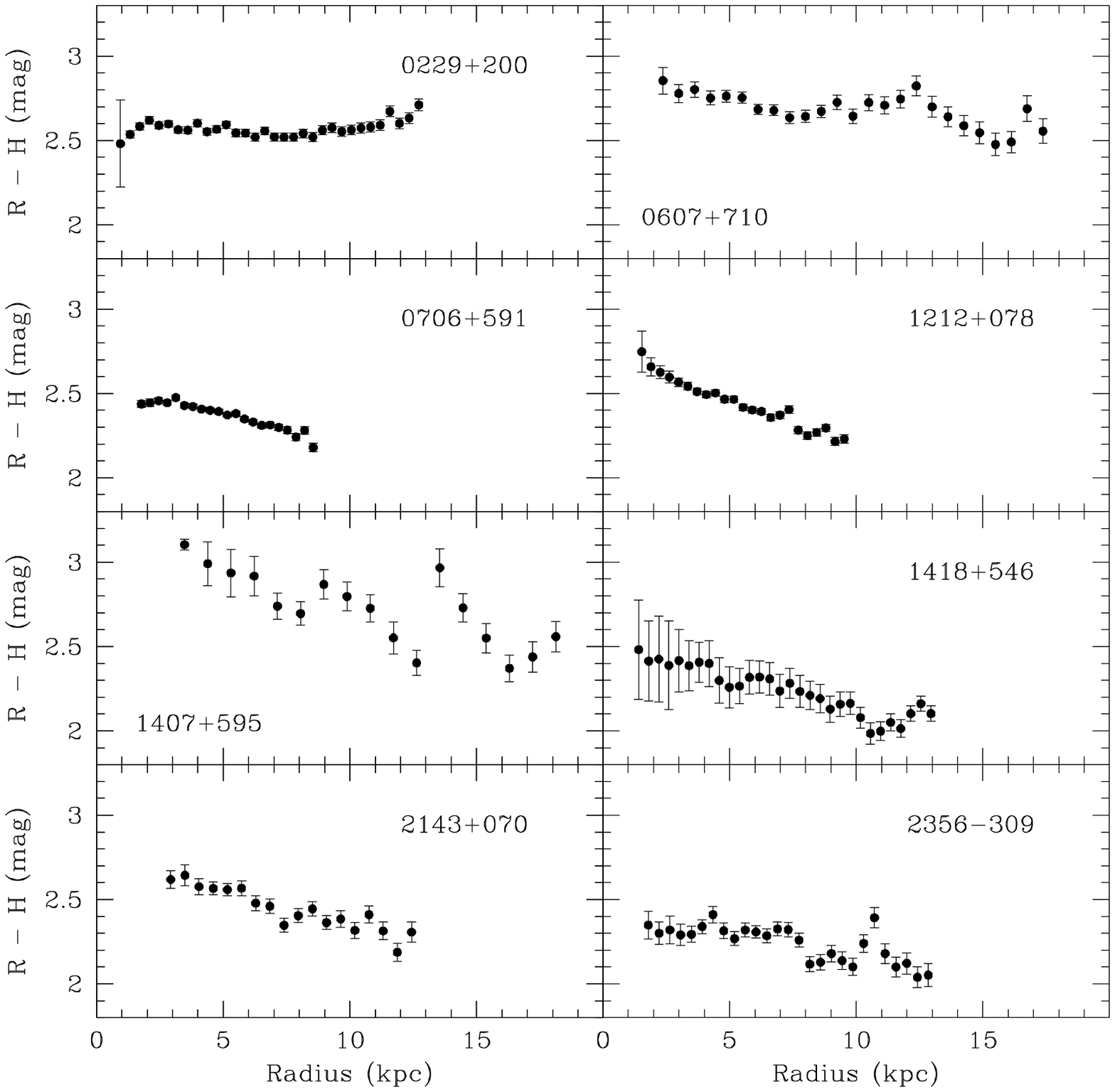,width=14cm}
\caption{
}
\end{figure}

\begin{figure}
\psfig{file=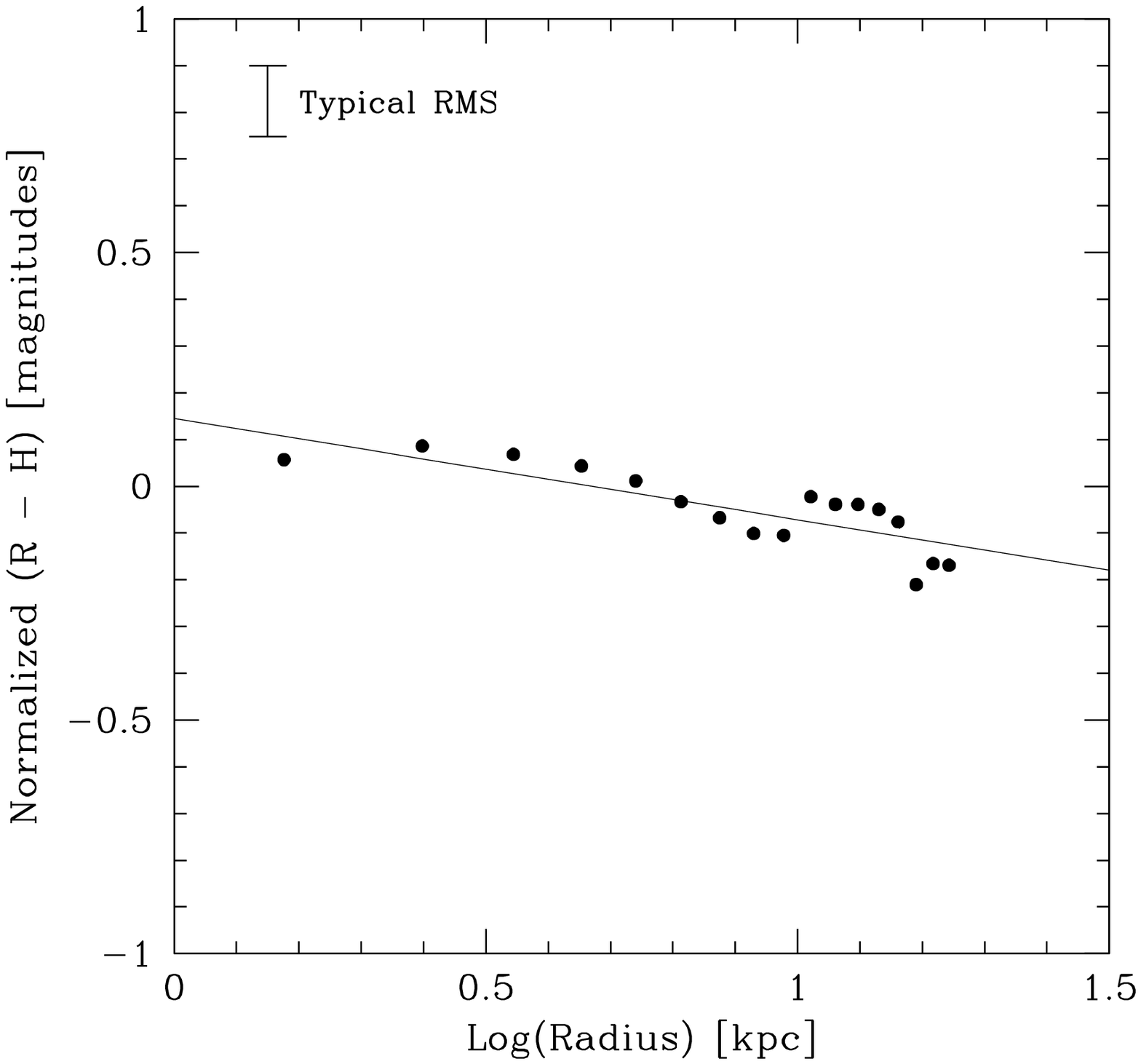,width=14cm}
\caption{
}
\end{figure}

\begin{figure}
\psfig{file=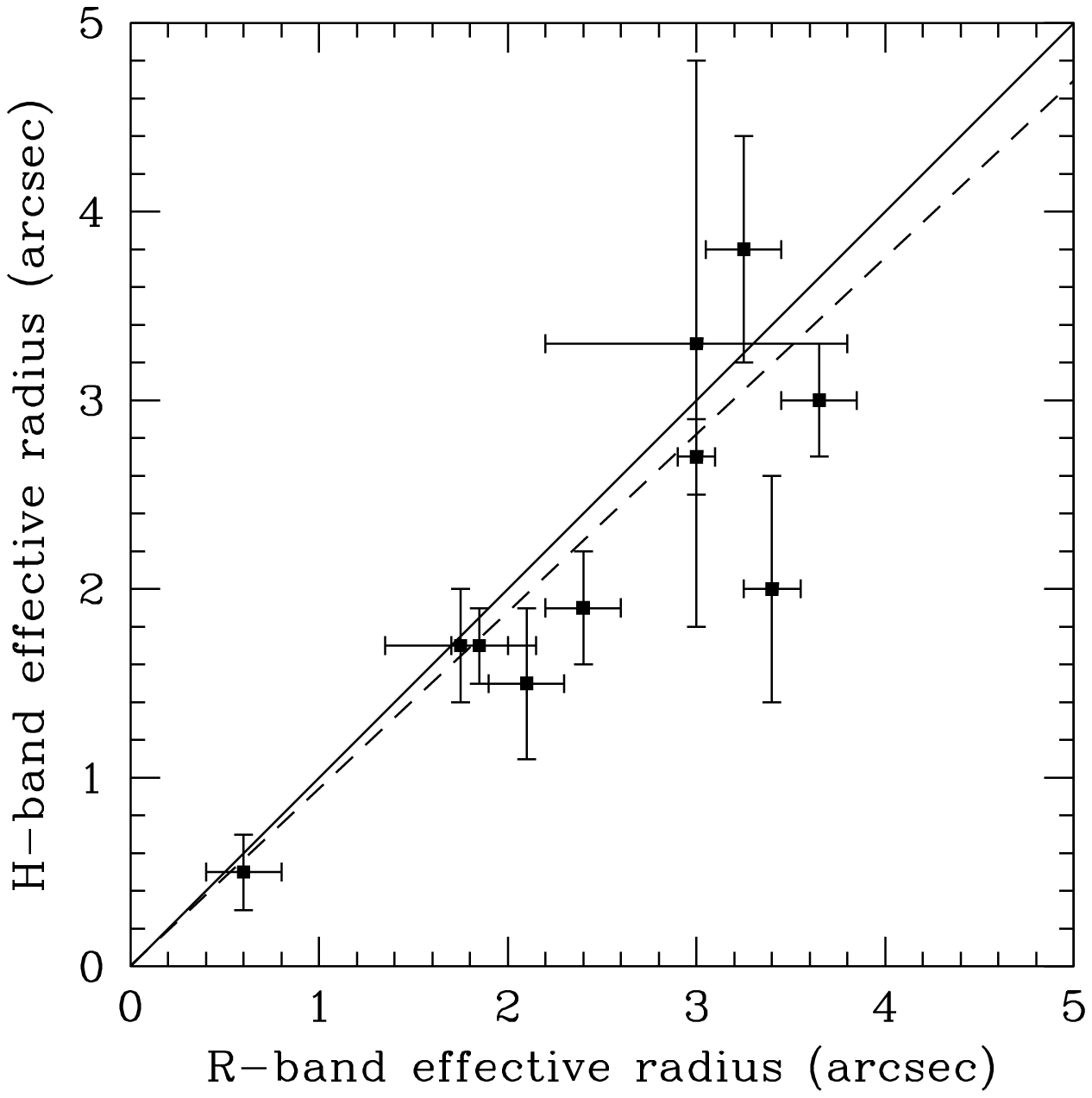,width=13cm}
\caption{
}
\end{figure}

\begin{figure}
\psfig{file=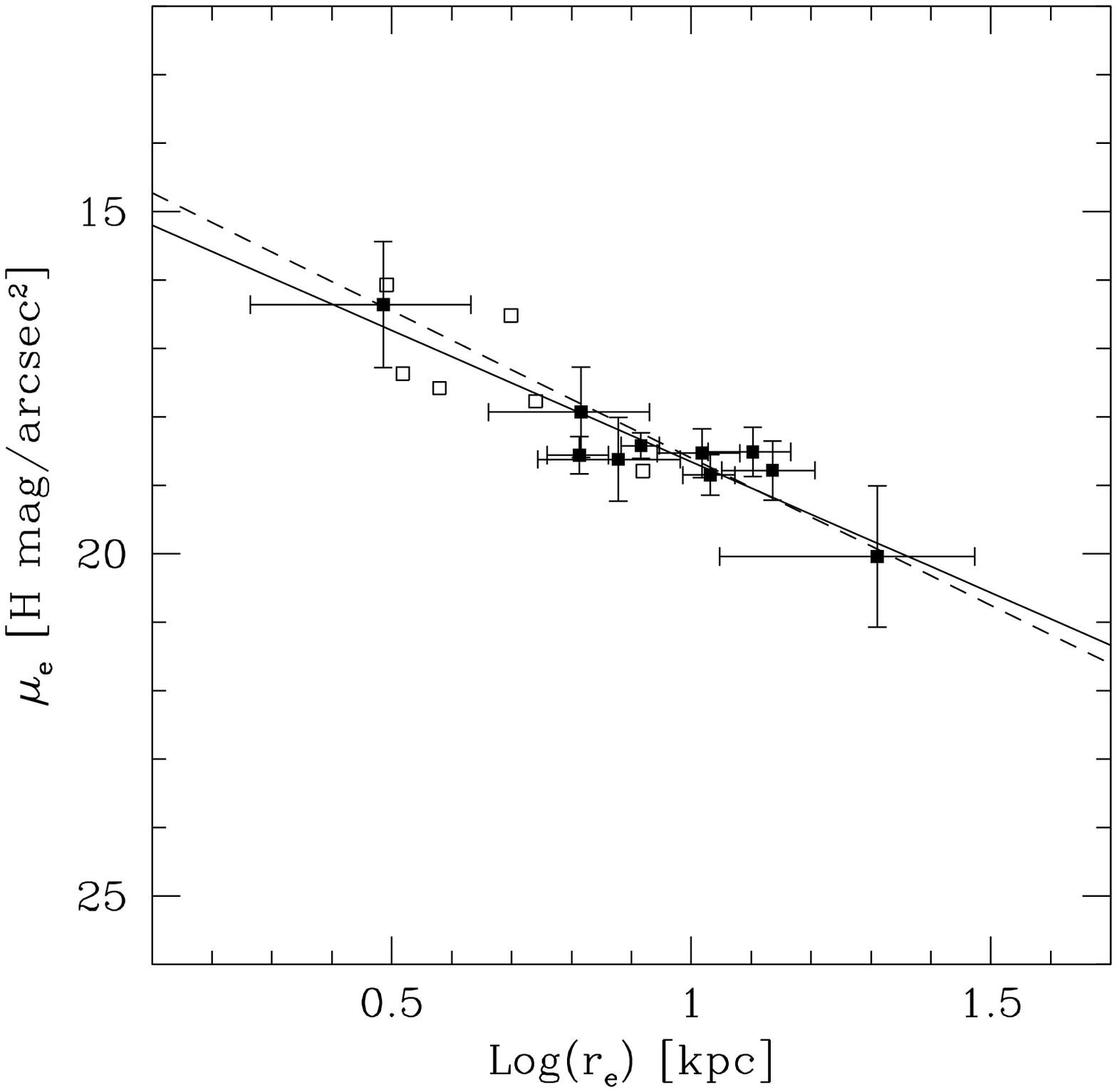,width=14cm}
\caption{
}
\end{figure}

\end{document}